
\input harvmac.tex
\input epsf.tex
\parindent=0pt
\parskip=5pt

\hyphenation{satisfying}

\def\IR{{\hbox{{\rm I}\kern-.2em\hbox{\rm R}}}}
\def\IB{{\hbox{{\rm I}\kern-.2em\hbox{\rm B}}}}
\def\IN{{\hbox{{\rm I}\kern-.2em\hbox{\rm N}}}}
\def\IC{\,\,{\hbox{{\rm I}\kern-.59em\hbox{\bf C}}}}
\def\IZ{{\hbox{{\rm Z}\kern-.4em\hbox{\rm Z}}}}
\def\IP{{\hbox{{\rm I}\kern-.2em\hbox{\rm P}}}}
\def\IH{{\hbox{{\rm I}\kern-.4em\hbox{\rm H}}}}
\def\ID{{\hbox{{\rm I}\kern-.2em\hbox{\rm D}}}}

\def\N{{\cal N}}

\noblackbox

\leftline{\epsfxsize1.0in\epsfbox{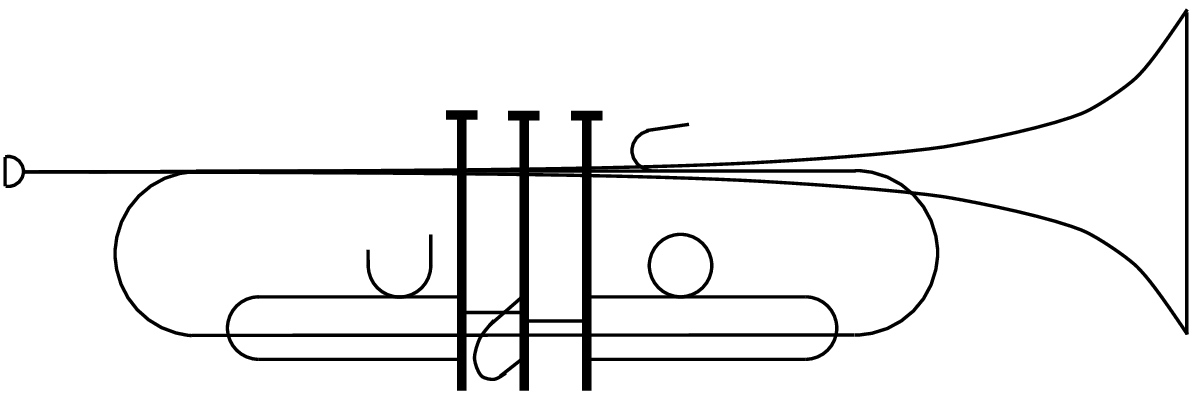}}
\vskip-0.9cm
\Title{\vbox{\baselineskip12pt
\hbox{UK/97--13}
\hbox{hep-th/9706155}}}
{From M--Theory to F--Theory, with Branes}

\centerline{\bf Clifford V. Johnson$^\dagger$}

\bigskip
\bigskip

\vbox{\baselineskip12pt\centerline{\hbox{\it Department of Physics and 
Astronomy}}
\centerline{\hbox{\it University of Kentucky}}
\centerline{\hbox{\it Lexington, KY 40506--0055 USA}}}
\footnote{}{\sl email: $^\dagger${\tt cvj@pa.uky.edu}}
\vskip1.7cm
\centerline{\bf Abstract}
\vskip1.7cm
\vbox{\narrower\baselineskip=12pt\noindent

A duality relationship between certain brane configurations in
type~IIA and type~IIB string theory is explored by exploiting the
geometrical origins of each theory in M--theory. The configurations
are dual ways of realising the non--perturbative dynamics of four
dimensional $\N{=}2$ supersymmetric
 $SU(2)$ gauge theory with four or fewer flavours
of fermions in the fundamental, and the spectral curve which 
organizes these dynamics plays
a prominent role in each case. This is an illustration of how
non--trivial F--theory backgrounds follow from M--theory ones,
hopefully demystifying somewhat the origins of the former.}


\Date{23rd June 1997}
\baselineskip13pt

\lref\dbranes{J.~Dai, R.~G.~Leigh and J.~Polchinski,  Mod.~Phys.~Lett.
{\bf A4} (1989) 2073\semi P.~Ho\u{r}ava, Phys. Lett. {\bf B231} (1989)
251\semi R.~G.~Leigh, Mod.~Phys.~Lett. {\bf A4} (1989) 2767\semi
J.~Polchinski, Phys.~Rev.~D50 (1994) 6041, hep-th/9407031.}
\lref\orientifolds{A. Sagnotti, in {\sl `Non--Perturbative Quantum
 Field Theory'}, Eds. G. Mack {\it et. al.} (Pergammon Press, 1988), p521\semi
V. Periwal, unpublished\semi J. Govaerts, Phys. Lett. {\bf B220}
(1989) 77\semi P. Hor\u{a}va, Nucl. Phys. {\bf B327} (1989) 461.}
\lref\nsfivebrane{A. Strominger, Nucl. Phys. {\bf B343}, (1990) 167; 
{\it Erratum: ibid.}, {\bf 353} (1991) 565.}
\lref\orientcft{S. Forste, D. Ghoshal and S. Panda, hep-th/9706057.}
\lref\townsend{P. Townsend, Phys. Lett. {\bf B150} (1995) 184, hep-th/9501068.}
\lref\towni{P.K. Townsend, Phys.Lett. B373 (1996) 68, hep-th/9512062.}
\lref\duffgibbtown{M. Duff, G. Gibbons and P. Townsend, Phys. lett {\bf B332}
 (1994) 321, hep-th/9405124.}
\lref\hanany{A. Hanany and E. Witten,  hep-th/9611230.}
\lref\elitzur{S. Elitzur, A. Giveon and D. Kutasov,  hep-th/9702014.}
\lref\usthree{N. Evans, C. V. Johnson and A. D. Shapere, hep-th/9703210, 
 to appear in Nucl.
 Phys.~{\bf B}.}
\lref\me{C. V. Johnson, hep-th/9705148.}
\lref\seibergwitten{N. Seiberg and E. Witten, 
Nucl. Phys. {\bf B426} (1994) 19; {\it Erratum: ibid.,}
 B430 (1994) 485, hep-th/9407087\semi N. Seiberg and E. Witten,
 Nucl. Phys. {\bf B431} (1994) 484, hep-th/9408099.}
\lref\wittentoo{E. Witten, hep-th/9703166.}
\lref\sen{A. Sen, Nucl. Phys. {\bf B475} (1996) 562, hep-th/9605150.}
\lref\banks{T. Banks, M. R. Douglas  and N. Seiberg, Phys. Lett.
 {\bf B387} (1996) 278, hep-th/9605199.}
\lref\edsmall{E. Witten, Nucl. Phys. {\bf B460} (1996) 541, hep-th/9511030.}
\lref\edbound{E. Witten, Nucl. Phys. {\bf B460} (1996)  335, hep-th/9510135.}
\lref\schwarzi{J. H. Schwarz, Phys. Lett. {\bf B360} 
 (1995) 13; {\it Erratum: ibid.}, {\bf B364} (1995) 252  hep-th/9508143.}
\lref\schwarzii{J. H. Schwarz,  Phys. Lett. {\bf B367} (1996) 97,
hep-th/9510086.}
\lref\mtwo{E. Bergshoeff, E. Sezgin and P. K. Townsend, Phys. 
Lett. {\bf B189} (1987) 75\semi M. J. Duff and K. S. Stelle,
Phys. Lett. {\bf B253} (1991) 113.}
\lref\mfive{R. G\"uven, Phys. Lett. {\bf B276} (1992) 49.}
\lref\double{M. J. Duff, P. S. Howe, T. Inami and K. S. Stelle, Phys. lett.
 {\bf B191} (1987) 70.}
\lref\aspinwall{P. Aspinwall, Nucl. Phys. Proc. Suppl. {\bf 46} (1996) 30,
 hep-th/9508154.}
\lref\town{P. Townsend, Phys. Lett. {\bf B350} (1995) 184, hep-th/9501068}
\lref\witt{E. Witten, Nucl. Phys. {\bf B443} (1995) hep-th/9503124.}
\lref\vafa{C. Vafa, Nucl. Phys. {\bf B469}  (1996) 403, hep-th/9602022.}
\lref\kkmonopoles{R. D. Sorkin, Phys. Rev. Lett. {\bf 51} (1983) 87\semi
D. J. Gross and M. J. Perry, Nucl. Phys. {\bf B226} (1983) 29.}
\lref\ericjoe{E. G. Gimon and J. Polchinski, Phys. Rev. {\bf D54} (1996) 
1667, hep-th/9601038.}
\lref\joeed{J. Polchinski and E. Witten, Nucl. Phys. {\bf B460} (1996) 525,
hep-th/9510169.}
\lref\klemm{A. Klemm, W. Lerche, P. Mayr, C. Vafa and N. Warner, Nucl. Phys. 
{\bf B477} (1996) 746, hep-th/9604034.}
\lref\katz{S. Katz, P. Mayr and  C. Vafa, hep-th/9706110.}


\newsec{\bf Introduction.}

\subsec{\sl Motivation}

Certain configurations of extended objects in string theory have
become of considerable interest of late, as they enable the intricate
interplay of duality, geometry, field theory and string theory to be
explored. Typically, these configurations involve
combinations\refs{\hanany,\elitzur}\ of D--branes\dbranes\ and
NS--(five)branes\nsfivebrane, and sometimes the
inclusion\refs{\usthree,\me}\ of orientifolds\orientifolds.
The field theories are realized in the dimensions common to all of the
world--volumes of the extended objects in question. The dynamics of
the field theories encode much of the geometrical behaviour of the
branes and {\it vice--versa}, yielding a powerful laboratory for the
study of familiar dualities and the discovery of new
ones.

These configurations are still somewhat novel, and many of their
properties remain to be fully understood. The aspects which we will
study in this paper are concerned with the question of how the physics
---as encoded in the world--volume field theory--- of a given
configuration can arise from a very different configuration of
extended objects. We are thus studying a sort of `dual pair' realizing
the same field theory, together with the properties of the
transformation between the members of the pair.

Consider for a moment the properties of `T--duality', acting on closed
string backgrounds. In the target geometry we can replace a circle of
radius $R$ by one of radius $\alpha^\prime/R$, where $\alpha^\prime$
is the inverse string tension. When the background fields have no
non--trivial dependence on the compact coordinate (at least
asymptotically), we understand what happens very well: winding and
momentum modes exchange roles, leaving the physics invariant. (Of
course, examining the action on space--time fermions, we see that the
type~IIA string theory is exchanged with the type~IIB.)

However in the open string sector, T--duality exchanges free boundary
conditions on the string endpoints with fixed ones (while exchanging
the circles), changing a D$p$--brane into a D$(p{+}1)$--brane or {\it
vice--versa}.  Therefore, T--duality applied to the multi--brane
configurations along\foot{Here, `T--duality along a direction' will
mean the following process in string theory: Compactify the direction
on a circle and shrink the radius of the circle to zero. When
T--duality applies, this process is equivalent to growing a new
non--compact direction, giving the `T--dual' configuration.}\ one of
the dimensions containing the field theory will change the dimension
of the field theory. This is {\sl not} the type of transformation
which we wish to consider. We wish to find a transformation on the
configuration which leaves the physical content of the field theory
invariant, including its dimensionality. As a result we must consider
transformations along a direction in which some branes are extended
and some branes are localized.

Necessarily therefore, we will study a transformation of
the brane configuration which is essentially a complicated version of 
T--duality. `Complicated' because it will involve two situations
where T--duality ---as phrased above--- is not well understood:
\item\item{\it (i)} It will
involve a direction along which the background fields (such as the
dilaton, metric and Kalb--Ramond field, all from the
Neveu--Schwarz/Neveu--Schwarz (NS--NS) sector) have non--trivial
dependence, because an NS--brane has its core there.
\item\item{\it (ii)} It will involve a direction along which the
world--volume of a D--brane is only of finite extent, because the
D--brane ends on the NS--brane. (This latter situation can be
interpreted as a non--trivial dependence of the Ramond--Ramond (R--R)
background fields on the coordinate in question.)

The end result of establishing the transformation will be a
realization of the {\sl same field theory} by either a brane
configuration in type~IIA string theory or a brane configuration in
type~IIB string theory.  As in each configuration the dilaton (and
hence the respective string couplings) varies from place to place in
space--time, it is more precise to say that we have a dual realization
involving M--theory and F--theory backgrounds.

\subsec{\sl Rephrasing T--Duality using M--Theory.}
The previous statement is the key to understanding just how we will
proceed.  In constructing the duality, we cannot use the strict
definition of T--duality given above at all stages, as it is tied very
much to the specific string theory context where the background field
dependence is relatively trivial. Note however, that for very simple
backgrounds we already know how we can embed our understanding of
T--duality between type~IIA and type~IIB string theory into a larger
context. First, recall that\refs{\town,\witt}: 
\item\item{\sl (a) Ten dimensional type~IIA string
theory is the zero radius limit of M--theory\foot{It will suffice in
the present context to take M--theory to mean `eleven dimensional
supergravity'. Strictly speaking, this is merely the low--energy limit
of M--theory, whatever it turns out to be.}\ compactified on a
circle.}

Placing
type~IIA on a circle and shrinking it to zero size, we have by
T--duality, an equivalent description in terms of ten dimensional
type~IIB string theory. The extra dimension is just the `T--dual'
dimension, which we understand very well in a stringy context as the
infinite radius circle dual to the one of zero radius upon which the
type~IIA theory is compactified.

Thinking of this two--step process as a single operation on M--theory
we arrive at the following conclusion\refs{\aspinwall,\schwarzi}: 
\item\item{\sl 
(b)  Ten dimensional type~IIB string theory is the zero size limit of
M--theory compactified on a torus.}

We will thus reinterpret T--duality between type~IIA and type~IIB
string theory as those statements about how to arrive at each theory
from M--theory.

\subsec{\sl Geometrical Origins of Branes and F--Theory.}
Nearly all of the D--branes in either theory have a simple
understanding in terms of the above geometrical statements ({\sl (a)}
and {\sl (b)}) together with the fact that M--theory contains two
basic branes, the M2--brane\mtwo\ and the M5--brane\mfive.

In type~IIA string theory, the D2--brane is a direct descendant of the
M2--brane, while the D4--brane is the double reduction of the
M5--brane\duffgibbtown, one dimension being wrapped on the circle.
The F1--brane ({\it i.e.,} the fundamental type IIA string) is the
double reduction of the M2--brane\double, while the F5--brane
(NS--brane) is the direct descendant of the M5--brane. The D0--brane
and D6--brane have a Kaluza--Klein origin as electric and magnetic
sources\refs{\town,\witt}.

Meanwhile, in the type~IIB string theory, the D1--brane and the
F1--brane come from wrapping one dimension of the M2--brane entirely
on one or other cycle of the $T^2$ \schwarzii.  Similarly, the
D5--brane and the F5--brane come from wrapping a dimension of the
M5--brane on one or the other cycle of the $T^2$. These partial
wrappings explain why the respective D-- and F--branes are mapped into
each other under the $\tau{\to}-1/\tau$ transformation of $T^2$ which
exchanges the two cycles.  Labelling them with integers (0,1) and
(1,0) respectively, the full $SL(2,\IZ)$ non--perturbative symmetry
produces a family of $(p,q)$ branes\refs{\schwarzi,\edbound}. The
D3--brane comes from wrapping two dimensions of the M5--brane on the
$T^2$, which explains\schwarzii\ why it is mapped to itself under
$SL(2,\IZ)$.

Understanding the existence of D7--branes in this geometrical picture
is the launching point for understanding the origins of
F--theory\vafa.  There, the configuration of seven--branes in the
non--perturbative type~IIB theory is given by the degeneration of an
auxiliary torus fibred over the ten physical dimensions of the theory.
The origin of this auxiliary torus is clear in the context of this
discussion.  Once we have arrived at the type~IIB string theory (using
{\sl (b)} above), we must not forget the torus upon which we
compactified M--theory.  We shrunk the area of the torus but we had a
choice about the complex tructure, $\tau$. Indeed, the type~IIB theory
`remembers' the complex structure of the torus, and this is frozen
into the resulting configuration. Im($\tau$) is
identified\refs{\schwarzi,\aspinwall,\vafa}\ with the inverse type~IIB
coupling $\lambda_{B}^{-1}{=}{\rm e}^{-\Phi}$, ($\Phi$ is the dilaton
field), while Re($\tau$) is the R--R scalar field $A^{(0)}$. The
degeneration of the auxiliary torus fibration is a jump in the value
of $A^{(0)}$, which signals the presence of a magnetic source of it, a
seven--brane. There is a $(p,q)$ family of these branes too, related
by $SL(2,\IZ)$, and the $(0,1)$ member of this family is the D7--brane
of perturbative type~IIB string theory.

We will take the position here that this is the geometrical origin of
F--theory: An elliptic fibration, defining a consistent type~IIB
background, is simply a concise way of specifying consistently a
collection of data about a {\sl family of tori} upon which M--theory
has been compactified before ultimately shrinking them away.

In M--theory, the D6--brane is a Kaluza--Klein monople\town, which
from a ten dimensional point of view is a circle fibration which
degenerates over the position of the D6--brane. This family of circles
becomes part of the family of tori which specify the data in
F--theory, as we will see. The degeneration of the circles (from the
ten dimensional point of view) ---signalling the presence of
D6--branes in type~IIA--- are inherited by the tori, ultimately
indicating the presence of D7--branes in type~IIB. We will also see
how other structures in type~IIA/M--theory give rise to some of the
other types of seven--brane of type~IIB/F--theory.  In this way, we
see that F--theory backgrounds are simply a subset of the possible
M--theory compactifications.

\subsec{\sl Beyond Simple T--Duality.}
So far, we have employed rather heavy machinery to carry out a task
which we can perform with simpler and sharper tools. We have recalled
the rephrasing of  T--duality and the taxonomy of branes in terms of the
geometry of M--theory. We already understand T--duality very well in
the terms laid out earlier, concerning the momentum and winding modes
of closed strings, and boundary conditions for open strings.

However, the simple geometric restating of T--duality reiterated here
is more readily adaptable to generalisation than the original
terminology. Indeed, we should be able to incorporate features which
we do not know how to handle well in the purely stringy context and we will
do so in what follows.

We can proceed to understand relationships between non--trivial brane
configurations in type~IIA and brane configurations in type~IIB as
follows: Interpret the type~IIA brane configuration as an M--theory
background. This renders harmless many features which are hard to
handle in string theory (such as branes ending on other branes) by
turning them into smooth M--theory
configurations\refs{\wittentoo,\usthree}. Next, compactify that
M--theory background upon a family of tori, chosen in a way which
respects the symmetries of the brane configuration, and shrink the
tori. The resulting background will be an F--theory background,
corresponding to a type~IIB configuration of extended objects with
non--trivial NS--NS and R--R background fields given by the data of
the shrunken tori.

Thus, the real use of the technique will become apparent when we try
to study the analogues of T--duality in directions where there is
non--trivial behaviour. The route described above will allow us
to realize an effective duality transformation which would have been
more difficult to determine using purely stringy techniques alone.

\subsec{\sl Outline}
The plan of this paper is as follows. In section~2, we will start by
describing the configuration of branes we wish to consider, in the
type~IIA string theory. It is essentially a review. Although it is a
classical discussion, it is a good starting point to orient ourselves,
and it will sometimes be useful to return to the classical
ten dimensional description for guidance.

In section~3, we review and follow the observation made in
refs.\refs{\wittentoo,\usthree}\ that to go beyond the classical
physics, it will be useful to go to a smooth description of the branes
as a configuration in M--theory, recovering within the brane geometry
the spectral curve\seibergwitten\ which controls the (Coulomb branch)
dynamics of the field theory\foot{It should be noted here that
  another type of situation
where the spectral curves of $\N{=}2$ field theories have been 
identified 
with the geometry of branes has been presented in the literature. 
(See ref.\klemm, for the original work and ref.\katz, for a recent 
extension.) This context of that work is somewhat different from the 
contexts of refs.\refs{\wittentoo,\usthree}\ and this paper,
 in which the identification is made after continuing to M--theory.}.

The detailed procedures for constructing such smooth descriptions were
presented in ref.\wittentoo, and we follow that presentation quite
closely, specializing to the case in hand, recovering the smooth
M--theory configuration as an M5--brane with topology
$\IR^4{\times}T^2$ in a multi--Taub--NUT geometry.

In section~4, we depart from what has gone before, walking the path
from M--theory to F--theory while carrying over the data of the
M5--brane/multi--Taub--NUT configuration. We arrive thus at section~5,
describing the F--theory configuration we expect to arrive at. Indeed,
the spectral curve for the field theory under consideration has been
previously recognized\sen\ as controlling the dynamics of a
seven--brane configuration in type~IIB/F--theory, and we make contact
with that description. It has also been pointed out\banks\ that the
$\N{=}2$, four dimensional field theory arises naturally on the
world--volume of a D3--brane probe moving around in the seven--brane
geometry. In our case, the D3--brane probe arises naturally as the
remains of the M5--brane we found in the M--theory: Its toroidal part
was wrapped on a space--time torus, which was subsequently shrunken
away.

In section~6 we discuss the type~IIB string theory ({\it i.e.},
classical) limit of the F--theory background, revisiting the work of
refs.\refs{\sen,\banks},  recognizing and interpreting certain
aspects of the `dual' type~IIA configuration in the new context.

We close with some remarks in section~7.

\newsec{\bf  The Type~IIA Brane Configuration.}

(This and the next section constitute a review ---tailored to our
needs--- and are included in order to set the scene, establish a few
conventions, and attempt a self--contained discussion.)

In this section the statements which we shall make will be essentially
classical ones, based on treating the fluctuations of flat branes.  We
will revisit this configuration in section~3, taking into account the
branes' deformations away from flatness caused by the forces they
exert on each other. As a result, the field theory content we will
deduce will be only true classically also.

Let us start with the following brane configuration in type~IIA string
theory:

\bigskip
\vbox{
$$\vbox{\offinterlineskip
\hrule height 1.1pt
\halign{&\vrule width 1.1pt#
&\strut\quad#\hfil\quad&
\vrule#
&\strut\quad#\hfil\quad&
\vrule width 1.1pt#
&\strut\quad#\hfil\quad&
\vrule#
&\strut\quad#\hfil\quad&
\vrule#
&\strut\quad#\hfil\quad&
\vrule#
&\strut\quad#\hfil\quad&
\vrule#
&\strut\quad#\hfil\quad&
\vrule#
&\strut\quad#\hfil\quad&
\vrule#
&\strut\quad#\hfil\quad&
\vrule#
&\strut\quad#\hfil\quad&
\vrule#
&\strut\quad#\hfil\quad&
\vrule#
&\strut\quad#\hfil\quad&
\vrule width 1.1pt#\cr
height3pt
&\omit&
&\omit&
&\omit&
&\omit&
&\omit&
&\omit&
&\omit&
&\omit&
&\omit&
&\omit&
&\omit&
&\omit&
\cr
&\hfil type&
&\hfil \#&
&\hfil $x^0$&
&\hfil $x^1$&
&\hfil $x^2$&
&\hfil $x^3$&
&\hfil $x^4$&
&\hfil $x^5$&
&\hfil $x^6$&
&\hfil $x^7$&
&\hfil $x^8$&
&\hfil $x^9$&
\cr
height3pt
&\omit&
&\omit&
&\omit&
&\omit&
&\omit&
&\omit&
&\omit&
&\omit&
&\omit&
&\omit&
&\omit&
&\omit&
\cr
\noalign{\hrule height 1.1pt}
height3pt
&\omit&
&\omit&
&\omit&
&\omit&
&\omit&
&\omit&
&\omit&
&\omit&
&\omit&
&\omit&
&\omit&
&\omit&
\cr
&\hfil NS&
&\hfil $2$&
&\hfil --- &
&\hfil --- &
&\hfil --- &
&\hfil --- &
&\hfil --- &
&\hfil --- &
&\hfil $\bullet$ &
&\hfil $\bullet$ &
&\hfil $\bullet$ &
&\hfil $\bullet$ &
\cr
height3pt
&\omit&
&\omit&
&\omit&
&\omit&
&\omit&
&\omit&
&\omit&
&\omit&
&\omit&
&\omit&
&\omit&
&\omit&
\cr
\noalign{\hrule}
height3pt
&\omit&
&\omit&
&\omit&
&\omit&
&\omit&
&\omit&
&\omit&
&\omit&
&\omit&
&\omit&
&\omit&
&\omit&
\cr
&\hfil D4&
&\hfil $2$&
&\hfil --- &
&\hfil --- &
&\hfil --- &
&\hfil --- &
&\hfil $\bullet$ &
&\hfil $\bullet$ &
&\hfil [---] &
&\hfil $\bullet$ &
&\hfil $\bullet$ &
&\hfil $\bullet$ &
\cr
height3pt
&\omit&
&\omit&
&\omit&
&\omit&
&\omit&
&\omit&
&\omit&
&\omit&
&\omit&
&\omit&
&\omit&
&\omit&
\cr
\noalign{\hrule}
height3pt
&\omit&
&\omit&
&\omit&
&\omit&
&\omit&
&\omit&
&\omit&
&\omit&
&\omit&
&\omit&
&\omit&
&\omit&
\cr
&\hfil D6&
&\hfil $N_f$&
&\hfil --- &
&\hfil --- &
&\hfil --- &
&\hfil --- &
&\hfil $\bullet$ &
&\hfil $\bullet$ &
&\hfil $\bullet$ &
&\hfil --- &
&\hfil --- &
&\hfil --- &
\cr
height3pt
&\omit&
&\omit&
&\omit&
&\omit&
&\omit&
&\omit&
&\omit&
&\omit&
&\omit&
&\omit&
&\omit&
&\omit&
\cr
}\hrule height 1.1pt
}
$$
}
\centerline{\sl Table 1.}

In  Table~1 (and in a similar one in section~6), a dash `---' represents
a direction {\sl along} a brane's world--volume while a dot~`$\bullet$'
is transverse to it. For the special case of the D4--branes' $x^6$
direction, where a world--volume is a finite interval, we use the symbol
`{\rm [---]}'.  (A `$\bullet$' and a `---' in the same column
indicates that one object is living inside the world--volume of the
other in that direction, and so they can't avoid one another. Two
`$\bullet$'s in the same column reveal that the objects are point--like
in that direction, and need not coincide in that direction, except for
the specific case where they share identical values of that
coordinate.)

In the configuration the D4--branes are stretched, in the $x^6$
direction, between the two NS--branes which are a distance
$x^6_1{-}x^6_2{=}L_6$ apart, where $x^6_{1,2}$ denote the positions of
the first and second NS--brane in the $x^6$ direction. The remaining
dimensions of their world--volumes, and that of all other branes, are
fully extended, filling the directions in which they lie.  

Consider the directions common to the world--volumes of all of the
branes.  There is a four dimensional field theory living on this
common space--time (with coordinates $(x^0,x^1,x^2,x^3)$). This field
theory has $\N{=}2$ supersymmetry, as the 32  supercharges are
reduced by half due to the presence of the NS--branes, 
and by a half again
due to the presence of the D4--branes. The presence of the
D6--branes does not break any more supersymmetries\hanany.

The (classical) field content of the four dimensional theory is easily
determined by the usual D--brane calculus: The excitations of open
strings stretching between the D4--branes (`4--4 strings') supply some
of the fields in the theory. Fluctuations parallel to the
world--volume supply a family of fields transforming as vectors under
the $SO(1,3)$ Lorentz symmetry. These vectors form $U(2)$ gauge bosons
(when the D4--branes are coincident).  Excitations transverse to the
world--volume represent the movement of the D4--branes. The D4--branes
must share the same position as the NS--branes in order to stay
tethered to  them, and therefore there are no fluctuations in the
$(x^6,x^7,x^8,x^9)$ directions. The only transverse fluctuations are
therefore in the $(x^4,x^5)$ directions which gives a set of complex
massless scalars in the field theory.  Taking into account their
transformation properties under the gauge symmetry, it is clear that
they form the complex adjoint scalar $\phi$, which lives in the
$\N{=}2$ vector multiplet. The strength of the gauge coupling $g$ is a
function of the distance between the NS--branes: $g^2\propto
\lambda_{A}/L_6$. Here, $\lambda_{A}$ is the type~IIA string
coupling, appearing in this way because the gauge kinetic term arises
in open string theory ({\it i.e.}, the D--brane sector) as a disc
amplitude.

The `matter' multiplets of the gauge theory are $N_f({\leq}4)$ families of
`quark': scalars in the fundamental of $U(2)$, which come from the
`6--4 strings' connecting the D6--branes to the D4--branes. The masses
of these quarks are set by the distance (in $(x^4,x^5)$) between the
D6--branes and the D4--branes.

The Higgs branch of the theory is reached by first making the quarks
massless by moving the D6--branes to be coincident with the
D4--branes. The D4--branes may now split, letting them have new
endpoints on the D6--branes, and the segments are now free to move
independently inside the D6--branes' world--volumes. The
$(x^7,x^8,x^9)$ positions parameterize the vacuum expectation values
(`vevs') of the quarks. In this way the gauge symmetry can be
completely Higgsed away.

The Coulomb branch of the theory (our concern for most of the paper)
is reached by giving the adjoint scalar $\phi$ a vev, with values in
the Abelian subalgebra of $U(2)$. This breaks the gauge symmetry down
to $U(1){\times}U(1)$ and corresponds to moving the D4--branes apart
in the $(x^4,x^5)$ directions. When a D4--brane encounters a D6--brane
in $(x^4,x^5)$, a quark becomes massless.

We need to understand this complicated brane configuration much
better. For example, the ending of the D4--branes on the NS--branes is
a somewhat singular situation. One might expect this feature to be
smoothed out in a way which corresponds to quantum corrections to the
field theory statements we have made in this section.  Ultimately, the
geometry reproduces the structure of the spectral
curves\seibergwitten\ which govern the structure of the quantum moduli
space of the gauge theories under discussion. This was anticipated and
exploited in ref.\usthree, and independently in ref.\wittentoo. In
ref.\wittentoo, the mechanisms by which the corrections to the brane
configurations may be deduced were explained, and the consequences
explored quite extensively.

\newsec{\bf The M--Theory Configuration.}

The starting point for correcting our classical configuration of the
previous section is to realize\wittentoo\ that the definite position
assigned the NS--branes in the $x^6$ direction is modified
considerably. The D4--branes, which are finite in that direction and
suspended between the NS--branes, are pulling the $(x^4,x^5)$ portion
of the NS--branes' world--volume out of shape, giving asymptotically
the shape of (say) the first NS--brane world--volume as:
\eqn\shapei{x^6_1=k\left(\ln|v-a_1|+\ln|v-a_2|\right)+{\rm const.},}
where $v{=}x^4{+}ix^5$, and $k$ is a constant which depends upon the
string coupling. Here, $a_1$ and $a_2$ are the positions of the two
D4--branes in the $(x^4,x^5)$ plane.

In order for the NS--brane's kinetic energy integral
\eqn\kinetic{\int\!\! d^4x\,
d^2v\sum_{\mu=0}^3 \partial_\mu x^6\partial^\mu x^6}
to converge, we have
\eqn\relation{a_1+a_2={\rm C},}
where C is some constant characteristic of the NS--brane. It can be
set to zero after a shift of the origin in $(x^4,x^5)$ space.  As
discussed before, the $a$ positions are the scalar components of the
gauge supermultiplet in the field theory. The sum $a_1{+}a_2$ controls
the overall $U(1)$ factor of the gauge group $U(2)$ and therefore
equation
\relation\ freezes out this $U(1)$, making our gauge group 
$SU(2)$. Considering the opposite D4--brane ends, on the other
NS--brane, leads to the same equation and no additional conditions on
the gauge group.

Turning to the gauge coupling, we revise our earlier formula to make
it a function of $v$:
\eqn\gaugei{{1\over g^2(v)}={x^6_1(v)-x^6_2(v)\over\lambda_{A}},}
and so we see that it is behaving as it should for a gauge theory,
varying as a function of some `mass scale' set by $|v|$: the quantity
$1/g^2$ diverges logarithmically as $|v|{\to}\infty$.

The next step is to recognize\refs{\usthree,\wittentoo}\ that this
type~IIA situation of D4--branes ending on and deforming NS--branes
should have a better description in M--theory. This is because on
going to M--theory an extra dimension unfolds, revealing that there
the D4--branes have a hidden world--volume dimension, and so become
M5--branes. The NS--branes also become M5--branes, with a definite
position in this new `M--direction', $x^{10}$. The parts of the
D4--branes we described in section~2 as lines in $x^6$ are actually
cylinders connecting the NS--branes. The final justification for going
to M--theory was pointed out in ref.\wittentoo: Looking at formula
\gaugei, it is clear that if we increase the string coupling
$\lambda_{A}$ while simultaneously increasing the inter--NS--brane
distance, the field theory is completely unaffected by
this. Therefore, we can go to the M--theory limit, where we grow an
extra dimension, $x^{10}$, of radius $R{\sim}\lambda_A^{2/3}$, as
measured in type~IIA units.

We now recognize\wittentoo\ that the formulae above were the real part
of a complex story. Giving the NS--branes positions in the $x^{10}$
direction, we have:
\eqn\shapeii{x^6_1+ix^{10}_1=R\left(\ln(v-a)+\ln(v+a)\right)+{\rm const.},}
and we may define the coupling (measuring now in M--theory units of
length\foot{Lengths measured in type~IIA units, $L_A$,
compare to lengths measured in M--theory units, $L_M$, by the formula
$L_A{=}R^{1/2}L_M$.})
\eqn\gaugeii{\tau(v)={\theta\over2\pi}+{4\pi i\over g^2(v)} 
={(x^{10}_1-x^{10}_2)\over 2\pi R} +
i{(x^6_1(v)-x^6_2(v))\over 2\pi R}.}  The angle $\theta$ changes
harmlessly by $\pm2\pi$ as an $x^{10}$ position of an NS--brane
changes by $2\pi R$, as it should.

We can quickly compute the $\beta$--function of our field theory using
the above formula as follows: Following the arguments of
ref.\hanany, made in the context of string theory
({\it i.e.}, the language of section~2), we know that we can move all
of the D6--branes past one of the NS--branes (let us choose the second
one), resulting in a D4--brane stretched from the NS--brane (starting
on the other $x^6$--side of it from the gauge D4--branes) to a
D6--brane, one for each D6--brane.

As the D6--branes are more massive than the D4--branes, 4--4 strings
entirely in the new D4--brane sector do not contribute to the gauge
group. However, the quarks are still present, as they now arise as
$N_f$ types of 4--4 string which connect the new D4--branes across the
NS--brane to the old D4--branes. Since the D4--branes on the other
side of the NS--brane pull the other way, the asymptotic shape of the
NS--brane with the extra branes is given by:
\eqn\shapeiii{x^6_2+ix^{10}_2=R\left(\sum_{i=1}^{N_f}\ln(v-m_i)
-\ln(v-a)-\ln(v+a)\right)+{\rm const.},} where the $m_i$ are the
D6--brane $(x^4,x^5)$ positions, or equivalently those of the new
D4--branes. They are the classical masses of the quarks.

Looking at the large $|v|$ behaviour of the coupling using this
formula, we get \eqn\betafunction{{2\pi i\tau(v)}=-(4-N_f)\ln v,}
displaying the one--loop $\beta$--function. When $N_f{=}4$ it
vanishes, as it ought to for the scale invariant theory.

The way\wittentoo\ to incorporate the D6--branes in this set--up
directly in the M--theory picture is to recognize\town\ that they
are Kaluza--Klein monopoles\kkmonopoles: The M--coordinate $x^{10}$ is
not simply a circle with which we form a product with the
$(x^4,x^5,x^6)$ directions to get the full space--time. Instead, it is
fibred over them in a Hopf--like fashion. The metric geometry of this
situation is that of multi--Taub--NUT. The positions of the D6--branes
are the positions in the base where the Killing vector for
translations in the $x^{10}$ circle vanishes, giving us a singularity
in the D6--brane metric when we reduce to ten dimensional type~IIA
string theory.

It is now clear that the type~IIA string theory configurations of
branes is a much less singular affair when viewed at strong coupling,
in M--theory. The D4--branes and NS--branes are just different
glimpses of the history of a single M5--brane's life--time.  If we add
a point representing infinity to the $(x^4,x^5)$ world--volumes of the
NS--branes, we see that in the full M--theory interpretation, the
world--volume of the M5--brane has topology $\IR^4{\times}T^2$, where
the $T^2$ is described as a surface embedded in the four dimensional
space~$Q_{N_f}$.  Here, $Q_{N_f}$ denotes the multi--Taub--NUT space
of multiplicity $N_f$, the M--theory origin of the $N_f$
D6--branes. In particular, $Q_0$ is just the product
$\IR^3{\times}S^1$ with coordinates $(x^4,x^5,x^6,x^{10})$. As pointed
out in ref.\wittentoo, it will suffice (for study of the Coulomb branch
of the field theories) to represent $Q_{N_f}$ as an equation of the
form:
\eqn\qnf{yz=\prod_{i=1}^{N_f}(v-m_i),} where $(y,z,v)$ are coordinates on
a three complex dimensional space with the structure of~$\IC^3$. As
before, $v{=}x^4{+}ix^5$. Defining the coordinate
$s{=}(x^6+ix^{10})/R$, we have that for fixed $z$, large $y$
corresponds to $t=\exp(-s)$ while for fixed $y$, large $z$ corresponds
to $t^{-1}$. The parameters $m_i$ are the $(x^4,x^5)$ positions of the
D6--branes.  We will require that the $N_f$ D6--branes are located {\sl
between} the NS--branes, and nowhere else. The specification \qnf\
misses (among other things) the $x^6$ positions of the D6--branes.

The world--volume of the M5--brane may be specified as a further
constraint equation in the coordinates $(y,v)$: $F(y,v){=}0$.
Giving $Q_{N_f}$ a complex structure and requiring holomorphicity in
$v$ and $y$ (very natural when viewed
from the point of view of the field theory) specifies the metric
structure on $T^2$ as a complex Reimann surface.

As a polynomial, the function $F$ must be quadratic in $y$ for a
($v{=}{\rm const.}$) slice to yield two  NS--branes in the
ten dimensional picture, and our constraint equation is thus of the
form\wittentoo:
\eqn\constraint{A(v)y^2+B(v)y+C(v)=0,} where $A,B$ and $C$ are
relatively prime polynomials.

There are no components of D4--branes extended outside the $x^6_1$ ---
$x^6_2$ interval; these would necessarily be semi--infinite (as they
have nothing else to end on), and as such would show up in our
solution as a divergence in $y$ for some definite value of $v$. The
absence of such behaviour fixes $A$ to be a constant, which we can
choose to be 1. The same requirement also removes the possibility of
$z$ diverging for some particular value of $v$ and this translates into
a condition on the form of $C$ and $B$ also: $C$ must have the same
zeros --with the same multiplicity\foot{This corresponds to placing
all of the D6--branes {\sl between} the NS--branes (in $x^6$), where
they can do some good, instead of outside the interval, where they are
irrelevant\wittentoo.}-- in the $v$ plane as has the defining
polynomial \qnf\ of $Q_{N_f}$, and $B$ must be quadratic in $v$ in
order to yield two D4--branes at fixed $y$ in the ten dimensional
picture.

Our torus is thus of the form:
\eqn\torus{y^2+B(v)y+f\prod_{i=1}^{N_f}(v-m_i)=0,}
where $f$ is an arbitrary complex constant. We can remove terms linear
in $v$ from $B(v)$ by a shift in $v$, which would shift the bare
masses $m_i$.  For the case $N_f{=}0$, the last term should simply be
a constant, which we can set to 1 without loss of generality. In terms
of ${\tilde y}{=}y{+}B/2$, we have:
\eqn\taurus{{\tilde y}^2={B(v)^2\over4}-f\prod_{i=1}^{N_f}(v-m_i)=0,} 
a standard form for the spectral curve
controlling the Coulomb branch of $\N{=}2$ supersymmetric four
dimensional $SU(2)$ gauge theory with $N_f$ quarks. The details of the
polynomial can be fixed by comparing to various field theory limits as
done in ref.\seibergwitten.

\newsec{\bf A New Direction.}

At the present stage, we have an M--theory background consisting of an
M5--brane with topology $\IR^4{\times}T^2$ propagating in the $N_f$
Taub--NUT space $Q_{N_f}$. The torus $T^2$ and the space $Q_{N_f}$,
are all described in terms of constraint equations in an auxiliary six
dimensional space.

Consider now the following. Let us ask instead for a slightly
different situation, which will differ from this one in ways which are
invisible in the field theory. Interpret the equation \constraint\ as
not only specifying the $T^2$ giving the shape of the M5--brane in the
four dimensional space $Q_{N_f}$, but also specifying two of the
space--time coordinates of the M--theory configuration. In other words,
{\sl we have wrapped the M5--brane we have been discussing on a
space--time torus of the same shape.} 

The manipulations following equation~\qnf\ and resulting in the final
curve \taurus\ serve to find us a smooth description of the wrapped
M5--brane on a space--time torus $T^2$, where the torus is fibred over
a base with topology~$\IR^2$. Some of the fibration data is inherited
from that of the multi--Taub--NUT geometry: The information about the
positions where the D6--branes live translates into a contribution to
the information about the location of zeros of the discriminant of the
torus fibration.

Let us return to the type~IIA description for a moment.  As the
Kaluza--Klein monoples feel no forces amongst themselves, it is not
problematic to have toroidally compactified one of the directions in
which they are point--like. The wrapping of the M5--brane on the torus
is already partially performed from the start: the D4--branes are a
piece of an M5--brane wrapped on the periodic $x^{10}$ direction. So
at any $x^6$ position where there is a D4--brane, we know that there
is a hidden part of an M5--brane wrapped on $x^{10}$.  What we have
effectively done is a further compactification of eleven dimensional
space--time. Focusing on the world--volume of an NS--brane, we must
make some combination of $(x^4,x^5)$ compact in order to get the
complete toroidal topology. We know from our experience with the
branes just how to do this: We simply add the space--time point at
infinity to the $(x^4,x^5)$ plane making it a~$\IP^1$, just as we did
to the world--volume of the NS--branes in those directions. The
$\IP^1$ has cuts or punctures in it due to the presence of the
D4--branes.

We have already seen\wittentoo\ that the size of the M--direction does
not affect the physics of the field theory if we rescale the
separation of the NS--branes accordingly. Similarly, the fact that we
have a $\IP^1$ for the $(x^4,x^5)$ direction (instead of $\IR^2$)
should not enter as a parameter in the field theory if we rescale the
positions of the D4-- and D6--branes to absorb any changes we make in
the overall size of the $\IP^1$.  

Returning to M--theory where the complete, smooth description is to
be found, we may now consider shrinking the $T^2$ part of the
M5--brane wrapped space--time. We hold the complex structure of the
torus (and hence the field theory data) fixed and shrink its size away
to zero.

\newsec{\bf  The  F--Theory Configuration.}

As described in the introduction, we know from simpler situations that
we have a type~IIB description of this situation (M--theory on a
shrunken torus) where:
\item\item{{\it (i)} 
We have a new  direction, $\hat x$, which restores us to a ten
 dimensional theory.}
\item\item{{\it (ii)} The wrapped M5--brane  becomes a D3--brane.}
\item\item{{\it (iii)} 
The data describing the shape of the torus which we shrink to zero
size is not lost, but is `remembered' by the final configuration: It
is frozen into an auxiliary torus, fibred over the ten dimensions of
the IIB theory. This is longhand for `F--theory'. }

As we know, the `data torus', or more precisely the family of such
tori, is that which specifies the Coulomb branch of the $\N{=}2$ four
dimensional $SU(2)$ gauge theory with $N_f$ quarks. Described as an
elliptic fibration over a base $\cal B$, with topology $\IR^2$, it is
singular over up to six points (depending upon $N_f$) in $\cal
B$. From the point of view of our F--theory background, these points
are the positions of magnetic sources of the R--R background field
$A^{(0)}$, as the modular parameter of the torus fibre specifies type~IIB
string background fields {\it via} the relation:
\eqn\relationiib{\tau(v)=A^{(0)}(v)+i{\rm e}^{-\Phi(v)},} where 
the type~IIB string coupling $\lambda_B(v)$ is related to the dilaton
$\Phi$ as $\lambda_B{=}{\rm e}^\Phi$. Such a magnetic source is an
object which is point--like in $\cal B$ and extended in the other
eight directions.  It is therefore a seven--brane of type~IIB
theory. In the case where we can describe the background with
perturbative type~IIB strings, the seven--brane is either a D7--brane
or an O7--plane (orientifold fixed plane). More generally, it can be
any of the infinite family of seven--branes which can appear in the
type~IIB theory by virtue of the $SL(2,\IZ)$ non--perturbative
symmetry.

The connection between precisely this family of tori \taurus\
(describing $D{=}4$, $\N{=}2$ $SU(2)$ gauge theory with $N_f$ quarks)
and an F--theory background was noticed in ref.\sen. It was pointed
out there that close to the perturbative type~IIB limit of F--theory
compactified on K3 ({\it i.e.,} the orbifold limit of the K3), the
background describes four identical families of six
seven--branes. Focusing on one family, in the 
limit two of the six possible singularities merge to become an
O7--plane while the rest become $N_f$ D7--branes. Furthermore, the
four dimensional field theory is naturally realized on the
world--volume of a D3--brane probe, as pointed out in ref.\banks. The
fact that the D3--brane has an $SU(2)$ living on it instead of just
$U(1)$ is T--dual to the fact\refs{\edsmall,\ericjoe} that it is
really {\sl two} D3--branes, plus an orientifold projection which
forces them to move together as a single object, projecting the
expected $U(2)$ (resulting from their coincidence) to $SU(2)$.

We see here that the D3--brane probe appears unbidden in this
framework as the wrapped M5--brane! We also know that the $N_f$
D7--branes have their origins in the presence of $N_f$ D6--branes,
while the O7--plane is an additional structure which was frozen
into the torus because of the non--trivial way (from the type~IIA
picture) the D4--branes end on the NS--branes. We can trace the
origins of the O7--planes to the D4/NS--brane system and not the
D6--branes because the case of no flavours has precisely two
O7--planes and no other singularities (not counting the point at
infinity).

In the next section we shall describe this further in the type~IIB limit.

\newsec{\bf  The Type~IIB Brane Configuration.}
 
Let us choose to label the coordinates of the base $\cal B$ by $v{=}{
x}^4{+}i{ x}^5$. (We should be careful here. This is not exactly the
$(x^4,x^5)$ pair of the type~IIA configuration.) Let us also denote by
${\hat x}^6$ the new, `dual direction' (which we briefly called $\hat
x$ in the last section).

We have the following  brane configuration in type~IIB string theory:

\bigskip
\vbox{
$$\vbox{\offinterlineskip
\hrule height 1.1pt
\halign{&\vrule width 1.1pt#
&\strut\quad#\hfil\quad&
\vrule#
&\strut\quad#\hfil\quad&
\vrule width 1.1pt#
&\strut\quad#\hfil\quad&
\vrule#
&\strut\quad#\hfil\quad&
\vrule#
&\strut\quad#\hfil\quad&
\vrule#
&\strut\quad#\hfil\quad&
\vrule#
&\strut\quad#\hfil\quad&
\vrule#
&\strut\quad#\hfil\quad&
\vrule#
&\strut\quad#\hfil\quad&
\vrule#
&\strut\quad#\hfil\quad&
\vrule#
&\strut\quad#\hfil\quad&
\vrule#
&\strut\quad#\hfil\quad&
\vrule width 1.1pt#\cr
height3pt
&\omit&
&\omit&
&\omit&
&\omit&
&\omit&
&\omit&
&\omit&
&\omit&
&\omit&
&\omit&
&\omit&
&\omit&
\cr
&\hfil type&
&\hfil \#&
&\hfil $x^0$&
&\hfil $x^1$&
&\hfil $x^2$&
&\hfil $x^3$&
&\hfil $x^4$&
&\hfil $x^5$&
&\hfil ${\hat x}^6$&
&\hfil $x^7$&
&\hfil $x^8$&
&\hfil $x^9$&
\cr
height3pt
&\omit&
&\omit&
&\omit&
&\omit&
&\omit&
&\omit&
&\omit&
&\omit&
&\omit&
&\omit&
&\omit&
&\omit&
\cr
\noalign{\hrule height 1.1pt}
height3pt
&\omit&
&\omit&
&\omit&
&\omit&
&\omit&
&\omit&
&\omit&
&\omit&
&\omit&
&\omit&
&\omit&
&\omit&
\cr
&\hfil O7&
&\hfil $2$&
&\hfil --- &
&\hfil --- &
&\hfil --- &
&\hfil --- &
&\hfil $\bullet$ &
&\hfil $\bullet$ &
&\hfil --- &
&\hfil --- &
&\hfil --- &
&\hfil --- &
\cr
height3pt
&\omit&
&\omit&
&\omit&
&\omit&
&\omit&
&\omit&
&\omit&
&\omit&
&\omit&
&\omit&
&\omit&
&\omit&
\cr
\noalign{\hrule}
height3pt
&\omit&
&\omit&
&\omit&
&\omit&
&\omit&
&\omit&
&\omit&
&\omit&
&\omit&
&\omit&
&\omit&
&\omit&
\cr
&\hfil D3&
&\hfil $2$&
&\hfil --- &
&\hfil --- &
&\hfil --- &
&\hfil --- &
&\hfil $\bullet$ &
&\hfil $\bullet$ &
&\hfil $\bullet$ &
&\hfil $\bullet$ &
&\hfil $\bullet$ &
&\hfil $\bullet$ &
\cr
height3pt
&\omit&
&\omit&
&\omit&
&\omit&
&\omit&
&\omit&
&\omit&
&\omit&
&\omit&
&\omit&
&\omit&
&\omit&
\cr
\noalign{\hrule}
height3pt
&\omit&
&\omit&
&\omit&
&\omit&
&\omit&
&\omit&
&\omit&
&\omit&
&\omit&
&\omit&
&\omit&
&\omit&
\cr
&\hfil D7&
&\hfil $N_f$&
&\hfil --- &
&\hfil --- &
&\hfil --- &
&\hfil --- &
&\hfil $\bullet$ &
&\hfil $\bullet$ &
&\hfil --- &
&\hfil --- &
&\hfil --- &
&\hfil --- &
\cr
height3pt
&\omit&
&\omit&
&\omit&
&\omit&
&\omit&
&\omit&
&\omit&
&\omit&
&\omit&
&\omit&
&\omit&
&\omit&
\cr
}\hrule height 1.1pt
}
$$
}
\centerline{\sl Table 2.}

Comparing Table~1 and Table~2, we see that from a string theory point
of view we have performed a sort of T--duality, in the $x^6$
direction. As one might expect, under it the D6--branes have turned
into D7--branes, as they should. Ignoring for a moment the finite
extent of the D4--branes in the $x^6$ direction, we see that they have
turned into a pair of D3--branes, as one might hope naively.  The
complication of the presence of the cores of two NS--branes, together
with the ending of a D4--brane on them, turns out to be `$T_6$--dual'
to an orientifold background. The orientifold procedure glues to the
two D3--branes into one dynamical object carrying an $SU(2)$ gauge
group, and introduces an O7--plane.

This perturbative type~IIB string background describes aspects of the
classical limit of the Coulomb branch of the $SU(2)$ gauge theory.
The position of the D3--brane in the $(x^4,x^5)$ plane parameterizes
the Coulomb branch of the gauge theory on its world volume, where the
gauge group is generically $U(1)$. As it moves around the plane, it
sees $N_f$ D7--branes each of charge 1 (in D7--brane units), and one
fixed plane, which is the O7--plane, the fixed plane of the
orientifold symmetry, which is $\Omega R_{45}$ on the bosonic
sector. If the D3--brane probe is coincident with the O7--plane, the
$SU(2)$ is restored. (Here, $\Omega$ is world--sheet parity, and
$R_{45}$ is $v{\to}{-}v$. The O7--plane has charge $-4$ as can be
deduced from requirements of $A^{(0)}$ charge cancellation in the full
compact situation: In total there are four O7--planes and sixteen
D7--branes.)

(As explained a while ago in ref.\banks, this is
understood in the $T_{45}$--dual type~I language as follows: The
D5--brane has gauge group $SU(2)$, resulting from a projection with
$\Omega$, in
constructing the type~I theory\refs{\edsmall,\ericjoe}. It has part of its
world--volume in the directions $(x^4,x^5)$ before doing the
$T_{45}$--duality to the present situation. This allows the
possibility of introducing $(x^4,x^5)$ Wilson lines (when making them
toroidal in preparation for the T--duality) to break the $SU(2)$ to
$U(1)$. These Wilson lines are $T_{45}$--dual to the positions of the
D3--brane probe here.)

Using the charge assignments just given, and the fact that the number
of transverse directions is two, one expects\sen\ that the couplings
are given by:
\eqn\couplings{\tau(v)= {1\over2\pi i}\left(\sum_{i=1}^{N_f}\ln(v-m_i)
-4\ln v\right)+{\rm const.},} where $m_i$ are the classical
positions of the D7--branes and we have placed  the
O7--plane at the origin. 

The similarity with the equations describing the asymptotic shape of
the NS--branes as they are pulled on by the D4--branes (in
section~3) should not escape our notice.  Combining equations
\shapeii\ and
\shapeiii, we have (placing the D4--branes at the origin):
\eqn\shapeiv{\eqalign{\tau(v)={\theta\over2\pi}+{4\pi i\over g^2}
&={(x^{10}_2-x^{10}_1)\over 2\pi R}+i{(x^6_1(v)-x^6_2(v))\over 2\pi
R}=\cr &= {1\over2\pi i}\left(\sum_{i=1}^{N_f}\ln(v-m_i)-4\ln v\right)
+{\rm const.}  }} The $m_i$ are the $(x^4,x^5)$ positions of the
D6--branes.  The similarity between the two formulae is not an
accident. It is part of the `dual' properties of the brane
configurations. Let us list some observations about these:

\item\item{\it (i)} In both cases there is $\N{=}2$ supersymmetry in 
four 
dimensions. The original thirty--two supercharges are reduced to
eight. In the type~IIA
case this is done by introducing NS--branes, and then D4--branes. 
Adding D6--branes to
the mix places no further constraints on the number of
supercharges. Similarly, in the type~IIB situation, there is a
$\IZ_2$  orientifold (which introduces an O7--plane),
followed by the introduction of a D3--brane. Adding D7--branes to
these does not `break' any more supersymmetry.

\item\item{\it (ii)}
In both cases, the logarithmic form of the two equations above is a
consequence of there being two relevant directions in which a
Laplace--Poisson equation is solved. In the type~IIA situation, it is
the two directions on the NS--brane in which the incident D4--branes
make a point, pulling in a transverse $x^6$ direction. In the type~IIB
scenario, it is the two directions transverse to both the
seven--branes and the D3--brane probe.
\item\item{\it (iii)} The main sources of non--trivial behaviour of the
dilaton in the type~IIA theory are the cores of the NS--branes, at
the place where the D4--branes meets them. Equation \shapeiv\ encodes
the asymptotic shape of the NS--branes' world--volumes, deformed in
the $x^6$ direction, and implicitly the distribution of background
NS--NS and R--R fields there. The `dual' configuration in type~IIB
makes this explicit: The D7--branes and O7--planes are NS--NS sources
for the dilaton and R--R sources for the field $A^{(0)}$, and equation
\couplings\ gives their asymptotic form, while the branes themselves
remain undeformed.
\item\item{\it (iv)} We can deduce that the  
D4/NS--brane system, non--trivial in the $(x^4,x^5,x^6)$ sector, acts
as an electric source for the R--R form $A^{(7)}$ in type~IIA, and hence
has some effective D6--brane charge, as measured by enclosing that
part of the configuration with a two--sphere at infinity. 

There are a number of ways to see that this is true: \item\item{ 
\hskip0.2cm
{\bf (a)} These charges are ultimately responsible for the O7--plane
(two extra seven--branes) in the `dual' type~IIB (F--theory) picture.
Interpreting our configurations as effectively $T_6$--dual to each
other, the O7--plane, carrying $A^{(0)}$ charge, is the  image under
$T_6$ of the D4/NS--brane junctions.}
\item\item{
\hskip0.2cm
{\bf (b)}  This charge assignment is consistent with
the fact that adding D6--branes, positioned precisely in the
$(x^4,x^5,x^6)$ directions, does not break any of the supersymmetries
already preserved by the D4/NS--brane configuration. From the point of
view of the D6--branes, adding them to the configuration is no
different from adding them to a system of parallel D6--branes.}
\item\item{
\hskip0.2cm
{\bf (c)} Possessing electric charge of  $A^{(7)}$ 
 is equivalent to having some magnetic $A^{(1)}$ (D0--brane)
charge. It is clear that the D4/NS--brane configuration has such
charge by considering the nature of the $x^6$ end--point of the
D4--brane in the $(x^4,x^5)$ part of the NS--brane's world--volume: It
is a `vortex' or monopole. As one circles a D4--brane's end--point
once in $(x^4,x^5,x^6)$ space\foot{It is appropriate to think of the
end--point as living in this space and not the $(x^4,x^5)$ space
alone once we take into account that the NS--brane is not at a
definite $x^6$ position.}\ and returns to the same position, some
winding has been acquired in the $x^{10}$ direction. This is the only
way to make local sense of the smoothing out of the D4/NS--brane IIA
system into a Reimann surface in M--theory\foot{I am grateful to  
M. Crescimanno and R. C. Myers for a  conversation apiece, concerning
this issue.}. This non--trivial winding is akin to the behaviour which we
attribute to a D6--brane in assigning it the role of a Kaluza--Klein
monopole of $A^{(1)}$.}
\item\item{\hskip0.2cm{\bf(d)} The $A^{(7)}$ 
charge observation is also consistent with the observation\hanany\
that moving a D6--brane through an NS--brane will result in a new
D4--brane stretched between them. Indeed, if we had moved the
D6--branes off to infinity, obtaining the quarks from the resulting
$N_f$ semi--infinite D4--branes instead, the final equation for the
shape of the M5--brane would have been precisely the same as the one
obtained here, \taurus\ \wittentoo. Hence, the F--theory result would
have been the same, and consequently so would be the final dual
type~IIB configuration in Table~2. Therefore, the effective $T_6$
duality treats the D4/NS--brane junction as an object with D6--brane
charge.}

\item\item{\it (v)} As pointed out in ref.\sen, the equation 
\couplings\ can only
be correct classically, or far away from the O7--plane. Close to the
orientifold, the imaginary part of $\tau$ would appear to be able to
go negative, which is not acceptable in a theory which is supposed to
be unitary. This is simply a reflection of the fact that there are
non--perturbative corrections to the formula \couplings\ as one
approaches the orientifold. The full solution is obtained by returning
to the complete F--theory background. The new non--perturbative data
are precisely those encoded in the spectral curve
\taurus, which yields the correct solution for $\tau$ everywhere and
hence the non--perturbative positions of the seven--branes. An
important fact is that the singularity at $v_0$, representing the
O7--plane, splits into two pieces, separated by a distance~${\rm
e^{\pi i\tau}}$. This corresponds to the O7--plane splitting into two
$(p,q)$ seven--branes in the full non--perturbative theory.
Similarly, the form \shapeiv\ for the shape of the NS--branes is only
true asymptotically; the complete data are in the M5--brane M--theory
configuration in the shape of the spectral curve \taurus.

Note that we can move from the theory with $N_f{=}4$
quarks to a lower number of quarks by the scaling limits described in
ref.\seibergwitten. For example, we send a D7-- (or D6--) brane
(corresponding to a a quark of mass $m$) off to infinity in the
$(x^4,x^5)$ plane. At the same time, we take the limit
$\tau{\to}i\infty$, and hold the product $\Lambda{=}{\rm e}^{\pi
i\tau}m$ fixed, defining the mass scale of the $N_f{<}4$ theory.

\newsec{\bf Closing Remarks.}

We have found that a type~IIA configuration of D4--branes,
NS--branes and D6--branes on whose intersection there lives an
$\N{=}2$ four dimensional $SU(2)$ gauge theory is related to a
type~IIB configuration of parallel D3--branes, D7--branes and
O7--planes, realizing the same gauge theory. The spectral curve
controlling the dynamics of the gauge theory appears naturally in the
topology and geometry of M--branes in M--theory on the one side, and
as F--theory data on the other.

We have studied a very non--trivial example of how F--theory brane
configurations may arise as M--theory ones, realizing an effective
T--duality in the process.

It seems that generalising the reverse process is always possible: We
should be able to start with an F--theory background and shrink a
direction over which the data torus is not varying much. This should
yield an M--theory background where the torus has now become
physical. If there were D3--branes in the F--theory background, they
will become M5--branes with two of their dimensions in the shape of
that torus. Returning to a type~IIA background by shrinking an
appropriate circle will yield a configuration of intersecting branes
of various sorts. This procedure should always be possibly locally,
and therefore we can understand (at least piece--wise) all F--theory
backgrounds in terms of M--theory brane configurations.

The generalisation of the M--theory to F--theory route (along the
lines of this paper) might be more challenging, however.  It would be
interesting to study how the example presented here might generalise,
providing a useful relation between certain type~IIA/M--theory brane
configurations and (pieces of) type~IIB/F--theory ones. There are many
reasons why this would be desirable. Much of the technology of
F--theory is very well organised in terms of the well--developed
geometry of elliptically fibred complex manifolds.  However, the study
of complicated M--theory/type~IIA brane configurations is still a
relatively new area, so being able to relate them to F--theory
backgrounds should help in sharpening certain aspects of their
analysis.

However, it is not clear that all relevant M--theory brane
configurations can be converted to F--theory ones in the specific way
done here.  Considering the case of higher rank gauge groups, where
the spectral curves are of higher genus than that of a torus, is
already interesting: the path to F--theory  will probably involve 
multiple wrappings of the M5--brane on the space--time torus, 
resulting in many  D3--branes in the final dual model, with 
additional discrete projections.

It will be interesting to study such issues further. The benefits of
finding a dictionary between M-- and F--theory configurations will be
of tremendous value in the study of the dynamics of gauge theories.

\bigskip
\medskip

\noindent
{\bf Acknowledgments:}

\noindent
CVJ was supported in part by family, friends and music. Thanks to 
E.~G.~Gimon and W.~Lerche for comments on the manuscript.

\bigskip
\bigskip

\centerline{\epsfxsize1.0in\epsfbox{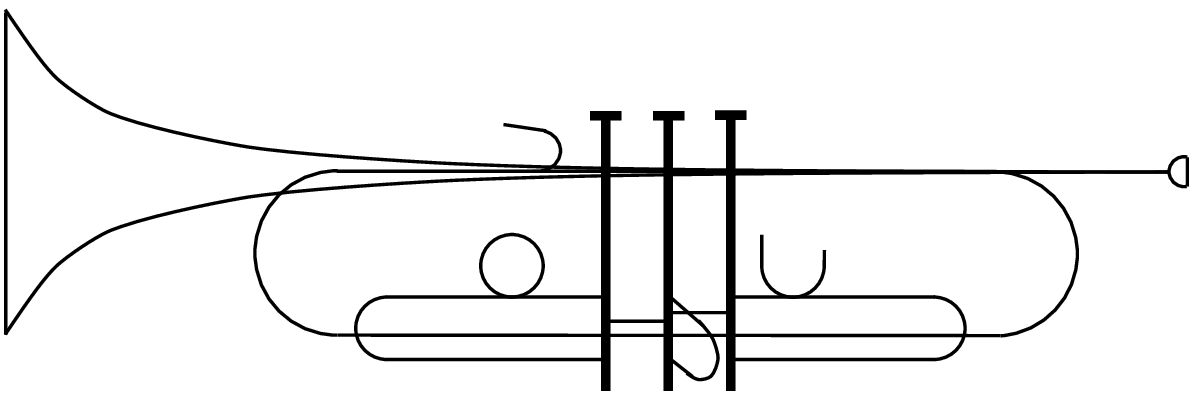}}

\listrefs

\bye